# Analysis of Indian foreign exchange markets: A Multifractal Detrended Fluctuation Analysis (MFDFA) approach

**Abstract:**

The multifractal spectra of daily foreign exchange rates for US dollar (USD), the British Pound (GBP), the Euro (Euro) and the Japanese Yen (Yen) with respect to the Indian Rupee are analysed for the period 6th January 1999 to 24th July 2018. We observe that the time series of logarithmic returns of all the four exchange rates exhibit features of multifractality. Next, we research the source of the observed multifractality. For this, we transform the return series in two ways: a) We randomly shuffle the original time series of logarithmic returns and b) We apply the process of phase randomisation on the unchanged series. Our results indicate in the case of the US dollar the source of multifractality is mainly the fat tail. For the British Pound and the Euro, we see the long-range correlations between the observations and the thick tails of the probability distribution give rise to the observed multifractal features, while in the case of the Japanese Yen, the origin of the multifractal nature of the return series is mostly due to the broad tail.

**Statements and Declarations**

No funding was received to assist with the preparation of this manuscript.

The authors have no competing interests to declare that are relevant to the content of this article

1. **Introduction**:

Financial systems are complex dynamical systems where many entities like traders, banks, brokerage firms, currency traders and mutual funds play a role by interacting with each other. The field has attracted physicists as well as economists, namely, Mantegna and Stanley (1999), Bouchaud and Potters (2004). Previous research in this has revealed that time-series data that has financial markets as its source reveal nonlinear properties. Some of the studies using methods from statistical physics were carried out by Mantegna et al. (1995, 1996), Plerou et al. (2003), Oh et al. (2006) and Qian et al. (2015).

Some of the properties that have been observed in financial time series are:

(a) long memory in volatility Liu et al. (1999), Yamasaki et al. (2005)

(b) multifractal nature Eisler (2004), Barunik et al. (2010), Kim (2004) and Calvet (2002)

(c) The probability distribution of logarithmic returns exhibit fat tails, Blattberg et al. (1974), Gopikrishnan et al. (2003).

The above characteristics are also known as "stylized facts". These properties of time series that have their origin in finance are found to be invariant universally across different markets, asset types and time scales. Power law kind of behaviour is observed by Pan et al. (2008) and Zhang et al. (2007) for the tails of the probability distribution function. These properties play an important role in risk mitigation strategies for investment in financial assets. Such similarities are also seen in critical phenomena and non-equilibrium behaviour and may be used to explain the data generating process underlying the economic time series model.

Many researchers have investigated the scaling behaviour of fractal models in financial markets. The idea of fractals in financial markets was introduced by Mandelbroit (1983). In further studies Mandelbroit (1974), McCaulay (1996) and Frisch (1996) have used the Multifractal analysis method to the intermittency in turbulent systems. The methodology has been applied to analyse time series of stock prices by Jiang et al. (2008,2009,2008,2007), interest rates time series by Cajueiro and Tabak (2007) and the time series of commodity prices by Matia et al. (2003) and He and Chen (2011). The efficiency of foreign exchange rates for developing countries like India have been studied by Datta and Bhattacharyya (2018a) using the Hurst exponent, and the predictability of the exchange rates have been reported in Datta and Bhattacharyya (2018b).

The multifractal model was used for analysing the foreign exchange rates bin the Asian markets by OH et al. (2012) . The variations in the exchange rate for the Iranian rial with the US dollar was examined by Norouzzadeh et al. (2006) considering their multifractal properties and scaling behaviour. The daily returns of the Shanghai stock price index were analysed by Yuan et al. (2009) using the multifractal model. Wang et al. (2011) studied the behaviour of the US dollar (USD) exchange rates. Evidence of multifractality was found by Ivanova and Ausloos (1999) using the 'structure-function method' in the Bulgarian Lev-USA Dollar (BGL-USD) exchange rate. Multifractality in Deutsche Mark/US Dollar exchange rates was reported by Schmitt et al. (2000). The multifractal properties of managed and independent float exchange rates were further investigated by Stosic et al. (2015). Kim et al. (2004) have reported evidence of multifractality for three financial assets, namely, the KOSPI, the won–dollar and yen-dollar exchange rates. Muniandy et al. (2001) have investigated the fractal nature and scaling behaviour of the forex rates for the Malaysian currency, Ringgit. In another study of relevance, the logarithmic returns from the two main financial indices (BSE and NSE) in India, was analysed using the MFDFA method by Kumar and Deo (2009). The US subprime crisis and its effect on the stock markets in US and Asia was the subject of study by Hasan and Mohammad (2015) for the time period 2007-2008. As far as we know there has been no specific research to look into the dynamics of the Indian foreign exchange rates using the MFDFA approach..

Different statistical properties of high frequency returns of the exchange rates of six currencies that form two triangles namely the EUR-GBP-USD and the GBP-CHF_JPY are studied by Drozdz et al. (2010 a).. In another paper Drozdz et al (2010 b) investigate the origins of multifractality in time series using the MFDFA method and on the Wavelet Transform Modulus Maxima (WTMM) methods. Gebarowski et al. (2019) used the Multifractal detrended cross-correlation methodology to study the foreign exchange (Forex) market time series. Kristoufek (2010) has reported on the properties of the Hurst exponent H particularly with regard to finite samples and its estimation using rescaled range analysis (R/S) and detrended fluctuation analysis (DFA) are described by Kristoufek (2010).

The objective of this paper is to examine the behaviour of four major world currencies which account for a large part of the forex transactions in India, namely, the US dollar (USD, the British Pound (GBP), the European Euro (Euro) and the Japanese Yen (Yen) exchange rates with respect to the Indian Rupee (INR). The method we use is the MFDFA (Multifractal Detrended Fluctuation Analysis) method and the time period of the study is from 6th January 1999 to 24th July 2018, a period of almost 20 years. There are two main causes that give rise to multifractal behaviour in any time series:(a) long term correlation (b) Broad tail in the probability distribution. To determine the source of multifractality in the Indian foreign exchange markets, we calculate the multifractal spectra first, using the original time series data. Then the data is shuffled randomly (which should remove the time correlations). Lastly, we create a surrogate time series using phase randomization on the original series (this should eliminate the effects due to the fat tails). The results are analysed, and studies are conducted to find the source of multifractality in the series under study.

The remaining part of this paper is organized in the following manner. The data used in the research are described in section 2. This is followed by section 3, where we describe the multifractal methodology in brief. Analysis of our results is presented in Section 4 and we conclude in Section 5.

## 2. Data

The daily closing prices of the four major foreign exchange rates with respect to the Indian rupee, namely the Us dollar (USD), the British Pound (GBP), the Euro (Euro) and the Japanese Yen (Yen) from 6th January 1999 to 24th J July 2018 were considered for the purpose of this study. This yields a total of 4726 observations. The closing index logarithmic returns, say Ret(n) = ln(price(n)−ln price(n−1) is considered for the calculations, where price(n) is the closing price on day n and price(n−1) is the closing price on day n−1. All the data are taken from the website of Reserve Bank of India, website tps://www.rbi.org.in/scripts/ReferenceRateArchive.aspx.

## 3 Methodology

Kantelhardt (2002) was the first to suggest the MFDFA method which is followed in this paper. The method is briefly outlined below.

Let $X_t (t = 1,2, ... , M)$ denote a time series of finite length $M$ where the number of points that have zero values is less or negligible compared to the non-zero values. If there is a value that is equal to zero i.e., $X_t = 0$, we interpret it as having no value at that instant. Another time series Y *(i)* is now calculated by accumulating the partial sums or integration as follows:

$$Y(i) = \sum_{t=1}^{i} [X(t) - <X>], \; i = 1,2, ... M, \qquad (1)$$

where the expression $<X>$ stands for the mean of $X_i$. The $Y(i)$ time series is further subdivided into $M_a =$ int $(M/a)$ segments of equal size $a$ which do not overlap with each other, The least square method is used to find the local propensity for each segment. In this procedure a short part of the time series Yi may be left out at the end. Since the number of points in the original time series $M$ is not necessarily exactly divisible by the length of the sub segment $a$, *we* replicate the same process starting from the other end of the time series. This procedure gives rise to a total of $2M_a$ segments. The local trend $y_\mu(i)$ within each of the $2M_a$ segments are now calculated using the least square fit, and the resulting variance is calculated as follows:

$$F^2(a,\mu) = \frac{1}{a}\sum_{i=1}^{a} \{Y[(\mu-1)a+i] - y_\mu(i)\}^2 \qquad (2)$$

for $\mu = 1,2, 3\ldots M$ and using

$$F^2(a,\mu) = \frac{1}{a}\sum_{i=1}^{a} \{Y[M-(\mu-M_a)a+i] - y_\mu(i)\}^2 \qquad (3)$$

for $\mu = N_a + 1, \ldots, 2N_a$. Here, the polynomial that is used to fit the data in box $\mu$ *is* denoted by $Y_\mu(i)$. It is well known that polynomials differ in their ability to eliminate the trends in the time series. This depends on the degree of the polynomial used to fit the data. Accordingly, we may use linear, quadratic, cubic or higher-order polynomials to appropriately fit the data.

Here the *p*th order fluctuation function is calculated as the average over all segments.

$$F_p(a) = \{\frac{1}{2M_a}\sum_{\mu=1}^{2M_a} [F^2(a,\mu)]^{\frac{p}{2}}\}^{\frac{1}{p}}. \qquad (4)$$

The value of the index variable *p* can be any real non-zero value. For *p*=0, we compute the fluctuation function as below:

$$F_p(a) = exp\{\frac{1}{4M_a}\sum_{\mu=1}^{2M_a} \ln[F^2(a,\mu)]\}. \qquad (5)$$

Finally, by analysing log $F_p(a)$ vs log *a* graphs for each value of *p* the scaling behaviour of the fluctuation function is determined.

If the variation $F_p(a)$ on the segment size *s* is similar to that of a power law then

$$F_p(a) \sim a^{h(p)} \qquad (6)$$

shows scaling in the nature of the fluctuation function. The notation h(q) also denoted as $H_q$, (as shown in the figures in the Results section), is called the generalized Hurst exponent in the standard literature. A special case for q=2 gives the standard well known Hurst exponent. For a monofractal time series, $H_q$ is independent of q, or in other words,

$$\Delta h(q) = Hq_{min} - Hq_{max} = 0 \qquad (7)$$

On the other hand, $H_q$ decreases with q for a multifractal time series. Hence $\Delta h(q) > 0$ is considered as a signature of multifractality in the time series. The relation between how complex a system is and how that is related to the multifractal nature of the system is characterized by $\Delta h_q$ as observed in a paper by Havlin et al. (1999). Zunino et al. (2009) have used the strength or degree of the multifractality present to relate to the inefficiencies present in financial systems.

The relation between the multifractal scaling exponent $H_q$ that can be determined from the MF-DFA calculations and the classical multifractal scaling exponent $\tau(q)$ or $\tau_q$ (denoted as $t_q$ in the figures) has been shown to follow the relationship depicted below by Kantelhardt (2002)

$$\tau_q = qH_q - 1 \quad (8)$$

The multifractal scaling exponent $\tau_q$ is also known as the q order Mass exponent or Renyi exponent. The Renyi exponent captures the time-varying properties of the time series and how it is related to the different moments. If $\tau q$ depends linearly on q, we may say that the time series is a monofractal, and if the Renyi exponent $\tau_q$ depends on q in a nonlinear manner, then the time series may be considered as a multifractal.

A separate way of analysing a multifractal spectrum of a time series is to study the singularity spectrum $D_q$ or $f(\alpha)$. It may be noted that here that $D_q$ or $f(\alpha)$ can be calculated by applying a Legendre transform on the variables q and $\tau_q$, as given below:

$$\alpha = \frac{d\tau q}{dq} \quad (9)$$

$$f(\alpha) = q\alpha(q) - \tau_q \quad (10)$$

In the above equation, the singularity strength is denoted by $\alpha$ also known as the Holder exponent. The singularities in a time series are characterized by the Holder exponent.

In the case of a time series that exhibits multifractal characteristics, the graph of $f(\alpha)$ shows a single hump which is converted to a delta function for a monofractal time series.

### 4. Results and Discussion

### 4.1 Multifractality in the return times series

The multifractal detrended fluctuation analysis (MFDFA) method is used in this study to look at the dynamics of the return time series of the four major exchange rates in India namely, the

US dollar (USD), the British Pound (GBP), the Euro (Euro) and the Japanese Yen (Yen) with respect to the Indian rupee. The time period considered is 6[th] January 1999 to 24[th] July 2018 yielding a total of 4726 points. The closing index logarithmic returns, that is, ln(price(t+1) −ln price(t) is considered for the calculations, the closing price on day t is denoted by price(t) and closing price on day t+1 is denoted by price(t+1).. In a previous paper (Datta et al. 2018) we have shown that the corrected empirical Hurst exponents calculated (using the method of Weron (2002 a, b) for the four exchange rates over the period 4[th] January 1999 to 11[th] April 2017 have the values 0.5865 for the USD, 0.4807 for GBP, 0.5080 for the Euro and 0.4714 for the Yen. These values show that the exchange rates, in general, show the presence of long memory and persistence/anti-persistence in their behaviour.

The MFDFA calculation results for USD are shown in Figures 1a, 1b, 1c and 1d. Fig1a shows the logarithm of the scaling function Fq (q-order RMS) for various scales (segment sample size) from 16 to 1024. Fig1b shows the q-order Hurst exponent Hq for various values of the moments q from -5 to +5. Fig 1c shows the q-order Mass exponent $\tau_q$ for q from -5 to +5 and Fig1d shows the multifractal spectrum Dq (also denoted by f(α) in the literature) versus hq.

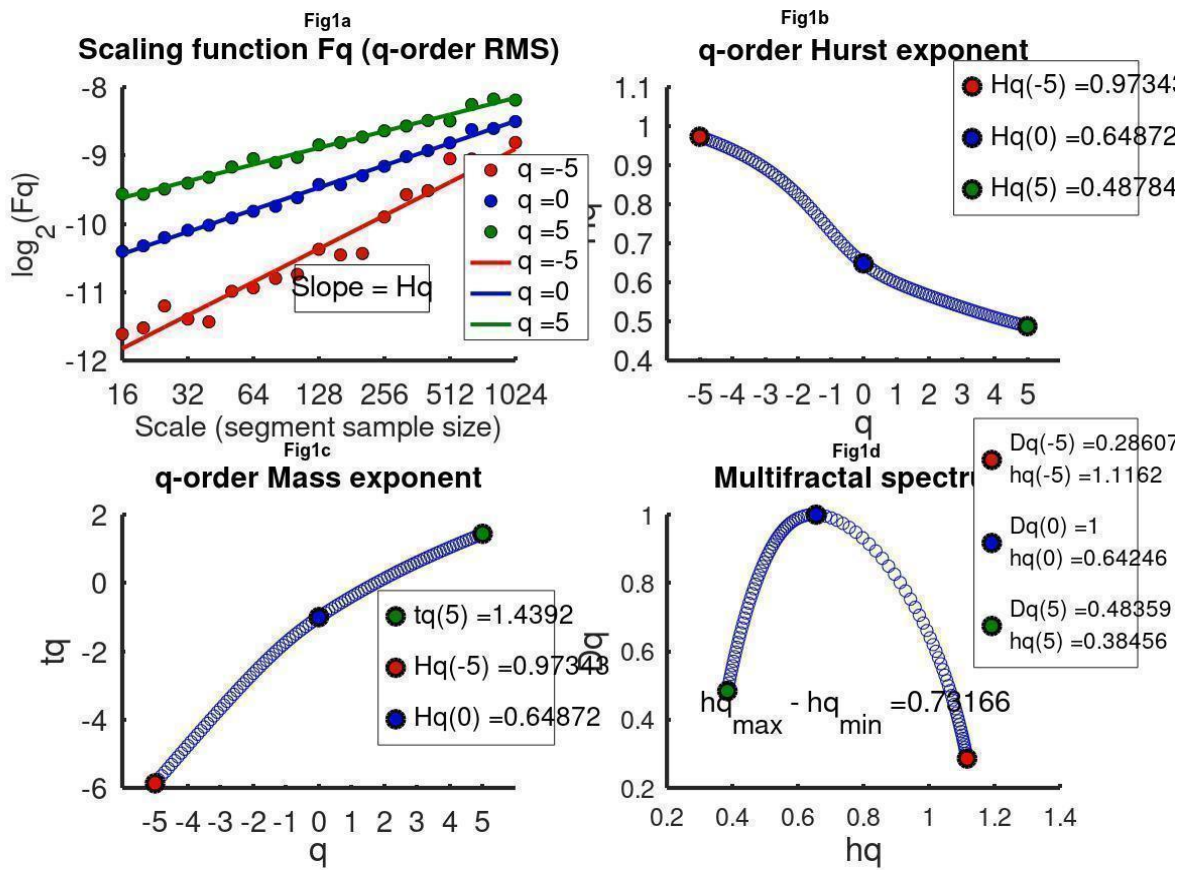

*Figure 1- For USD a) logarithm of the scaling function Fq (q-order RMS) for various scales (segment sample size) from 16 to 1024. (b)The q-order Hurst exponent Hq for various values of the moments q from -5 to +5. (c)The q-order Mass exponent Tq for q from -5 to +5 (d) The multifractal spectrum Dq versus hq*

The same results for GBP are presented in Figures 2a,2b,2c and 2d, the results for the Euro are presented in Figures 3a,3b,3c and 3d and the results for the Yen are presented in Figures 4a, 4b, 4c and 4d.

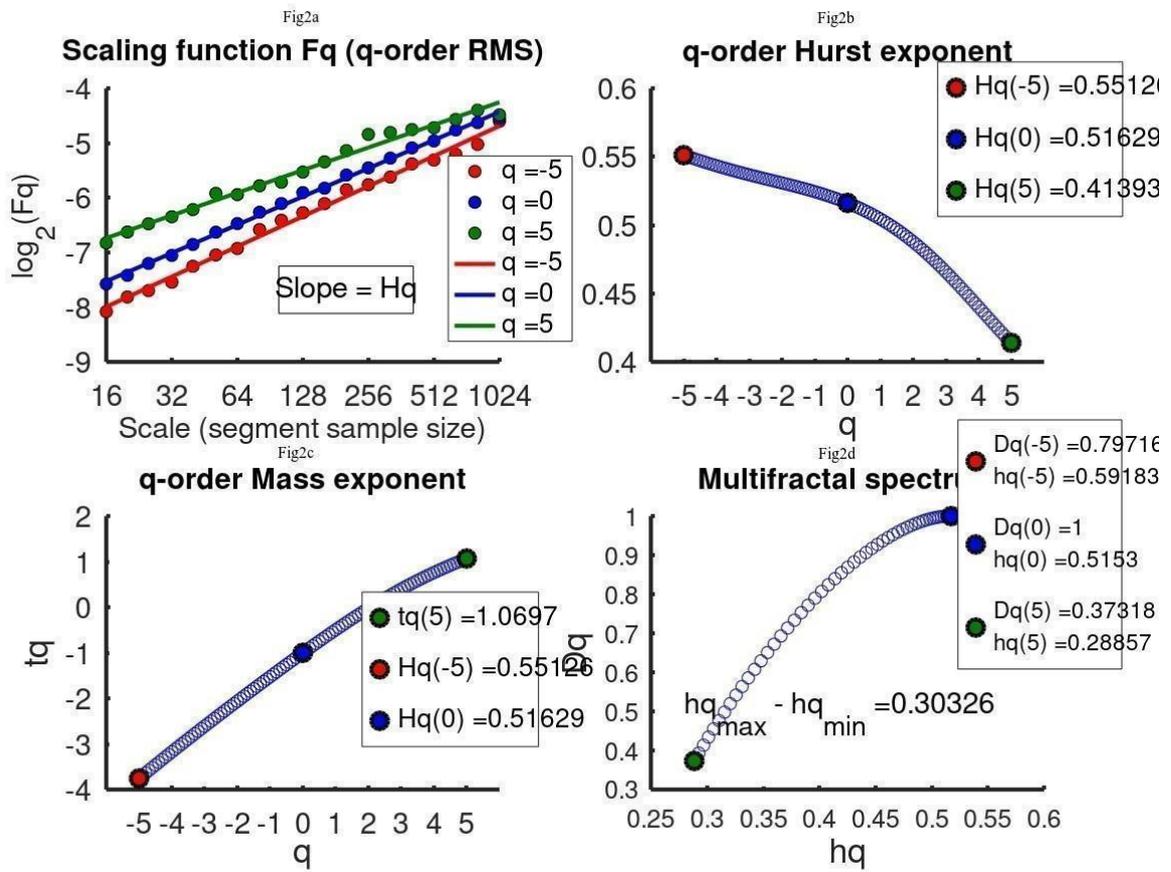

*Figure 2: GBP (a) logarithm of the scaling function Fq (q-order RMS) for various scales (segment sample size) from 16 to 1024. (b)The q-order Hurst exponent Hq for various values of the moments q from -5 to +5. (c)The q-order Mass exponent $\tau_q$ for q from -5 to +5 (d) The multifractal spectrum Dq versus hq*

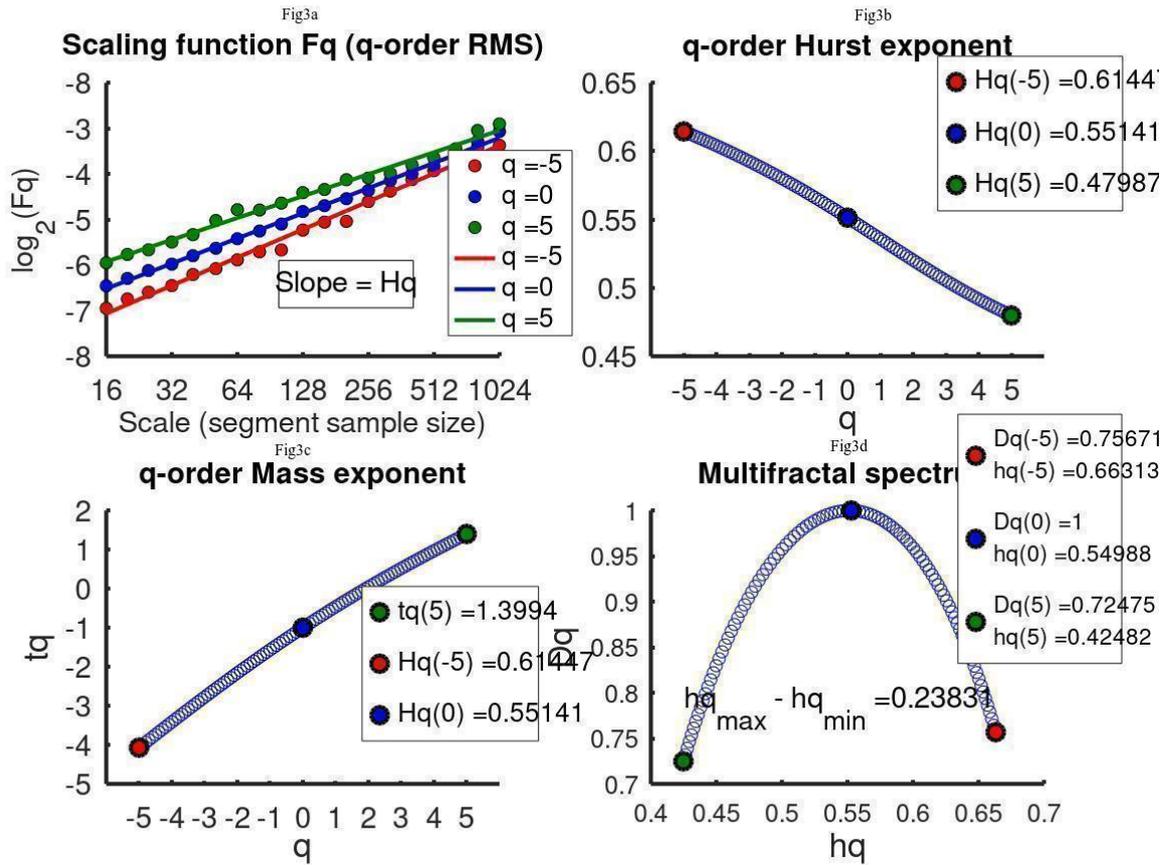

*Figure 3: Euro (a) logarithm of the scaling function Fq (q-order RMS) for various scales (segment sample size) from 16 to 1024. (b)The q-order Hurst exponent Hq for various values of the moments q from -5 to +5. (c)The q-order Mass exponent $\tau_q$ for q from -5 to +5 (d) The multifractal spectrum Dq versus hq*

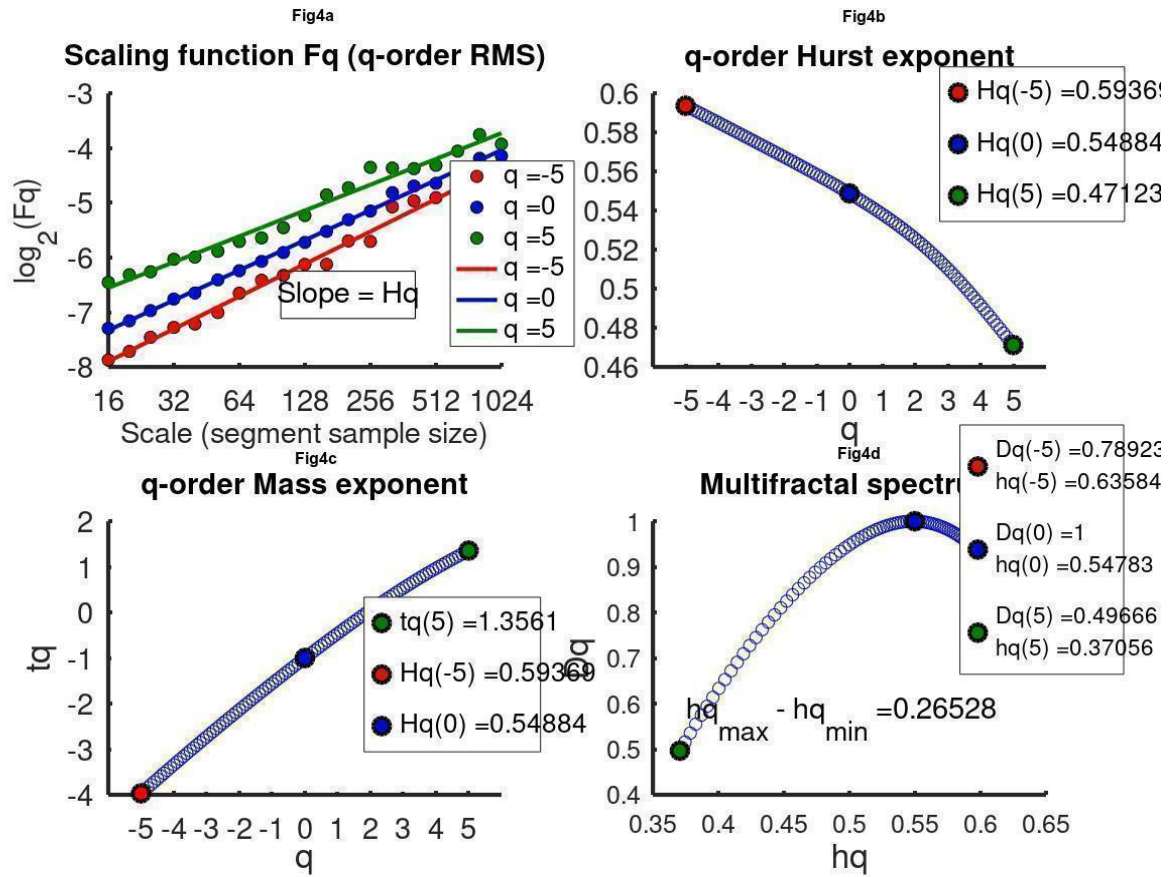

*Figure 4: Yen (a) logarithm of the scaling function Fq (q-order RMS) for various scales (segment sample size) from 16 to 1024. (b)The q-order Hurst exponent Hq for various values of the moments q from -5 to +5. (c)The q-order Mass exponent $\tau_q$ for q from -5 to +5 (d) The multifractal spectrum Dq versus hq*

The impact of fluctuations that are fast-changing will influence the overall scaling function Fq for those segments that have a small sample size or small scale for the return time series of USD, GBP, Euro and Yen. Likewise, the fluctuations that change slowly will impact the overall Fq for those segments that have a large sample size or are large scale. The scaling function, Fq, shown in Fig1a,2a,3a and 4a is therefore calculated for multiple segments sizes (i.e., multiple scales). This is to stress the fact that both fast and slow evolving fluctuations influence the structure of the time series. For the sake of clarity, we only show the graphs for q=-5,0 and +5 in the figures. It may be noted that Hq can be calculated from the slope of Figures 1a,2a,3a and 4a, respectively. From the figures, we can see that Hq changes its value depending on q, with the highest change seen for the USD

From figures 1b,1c, 2b,2c,3b,3c,4b,4c we see that the scaling exponents Hq and Tq for the returns of the USD, GBP, Euro and Yen show nonlinear dependence. The concavity in the Tq plots shown in figures 1c,2c,3c and 4c for all original return series also show the presence of multifractality. This also points towards the presence of multiple singularity exponents. This proves that the return series for these four exchange rates with respect to the Indian rupee exhibit multifractal behaviour. Another signature of multifractality is the shape of the Multifractal spectrum which is the plot of Dq versus hq shown in Figures 1d,2d,3d and 4d for the USD, GBP, Euro and Yen, respectively. For a multifractal structure, the shape of the multifractal spectrum should resemble that of an inverted parabola. Figures 1d,2d,3d and 4d corroborate that the shape does resemble that of an inverted parabola. We can define the strength of the multifractality ($\Delta\alpha$) as follows:

$\Delta\alpha = \alpha_{max} - \alpha_{min}$

This can also be expressed as

$\Delta hq = hq_{max} - hq_{min}$

in the notation used in Figures 1d,2d,3d, and 4d.

A higher value of $\Delta\alpha$ points towards higher degree or strength of multifractality which in turn may be related to higher complexity of the system under consideration.

We observe that the USD return series is characterized by the highest strength of multifractality $\Delta\alpha = 0.73166$ while the Euro return series has the lowest strength $\Delta\alpha = .23831$. The $\Delta\alpha$ values for the GBP and Yen are 0.30326 and 0.26528, respectively, which lie in between USD and Euro. Linking this with the theory of market efficiency we may say that the USD shows the highest complexity, i.e., least market efficiency while the Euro return series shows the least complexity and hence highest market efficiency among the four exchange rates considered in this paper. The market efficiency of the GBP and Yen lie in between these two exchange rates.

**4.2 Sources or Origins of Multifractality**

Kantelhardt et al. [2002] observed that there are two different origins of multifractality in a time series: (i) The long-range time-dependent correlations for small and large fluctuations, and (ii) Multifractality that is characterized by the wide tails of the probability distributions of the variations. Observations over multiple scaling exponents are needed to study the small and large fluctuations for both these sources. To find which source contributes how to the overall multifractality we make use of two transformations on the original return series namely: (a) shuffling the data and (b) phase randomization.

By shuffling the return series, the distribution of the various moments is kept intact but the long-range correlations are removed. After random shuffling, the data has the same distribution but without any time correlations or memory. Each time series in our calculations has been shuffled 100 times, and the calculations for the scaling function Fq, the q-order Hurst exponent Hq, the q order Mass exponent function Tq, and Multifractal spectrum are repeated for the shuffled time series.

The question arises as to what effect of the broad tails of the distribution function have on the multifractal nature of the logarithmic returns obtained from the various exchange rate time series data. To study this the method of phase randomization is used in this paper. The phase randomization process creates surrogate data blocks but preserves the second-order properties of the original time series data set. The original data is converted into the frequency domain, randomizing the phases simultaneously across the time series and converting the data back into the time domain (Prichard and Theiler 1994). The steps involved in phase randomization are as follows:

(i) In the first step the data points in the time series are subjected to the discrete Fourier transformation
(ii) In step two, the transformed data from step (i) is multiplied with random phases.

(iii) Finally in step three we calculate the inverse Fourier transform on the output of step two to obtain the phase-randomized times series.

The MFDFA calculation results for USD, GBP, Euro and Yen for the shuffled return series are shown in Figures 5, 6,7 and 8. Similarly, the MFDFA calculation results for USD, GBP, Euro and Yen for the phase randomized return series are presented in Figures 9,10,11 and 12. Figures 5a, 6a,7a and 8a show the logarithm of the scaling function Fq (q-order RMS) for various scales (segment sample size) from 16 to 1024 for the shuffled USD, GBP, Euro and Yen return series, respectively. Fig5b, 6b,7b and 8b display the q-order Hurst exponent Hq for different values of the moments q from -5 to +5 for the shuffled return series of the USD, GBP, Euro and Yen, respectively. Figures 5c, 6c, 7c and 8c show the q-order Mass exponent Tq for q from -5 to +5 for the shuffled USD, GBP, Euro and Yen return series, respectively. Figures 5d,6d,7d and 8d show the multifractal spectrum Dq (also denoted by f(α) in the literature) versus hq for the shuffled return series of USD, GBP, Euro and Yen, respectively.

The same results for the phase randomized return series for USD, GBP, Euro and Yen respectively are presented in Figures 9,10,11 and 12.

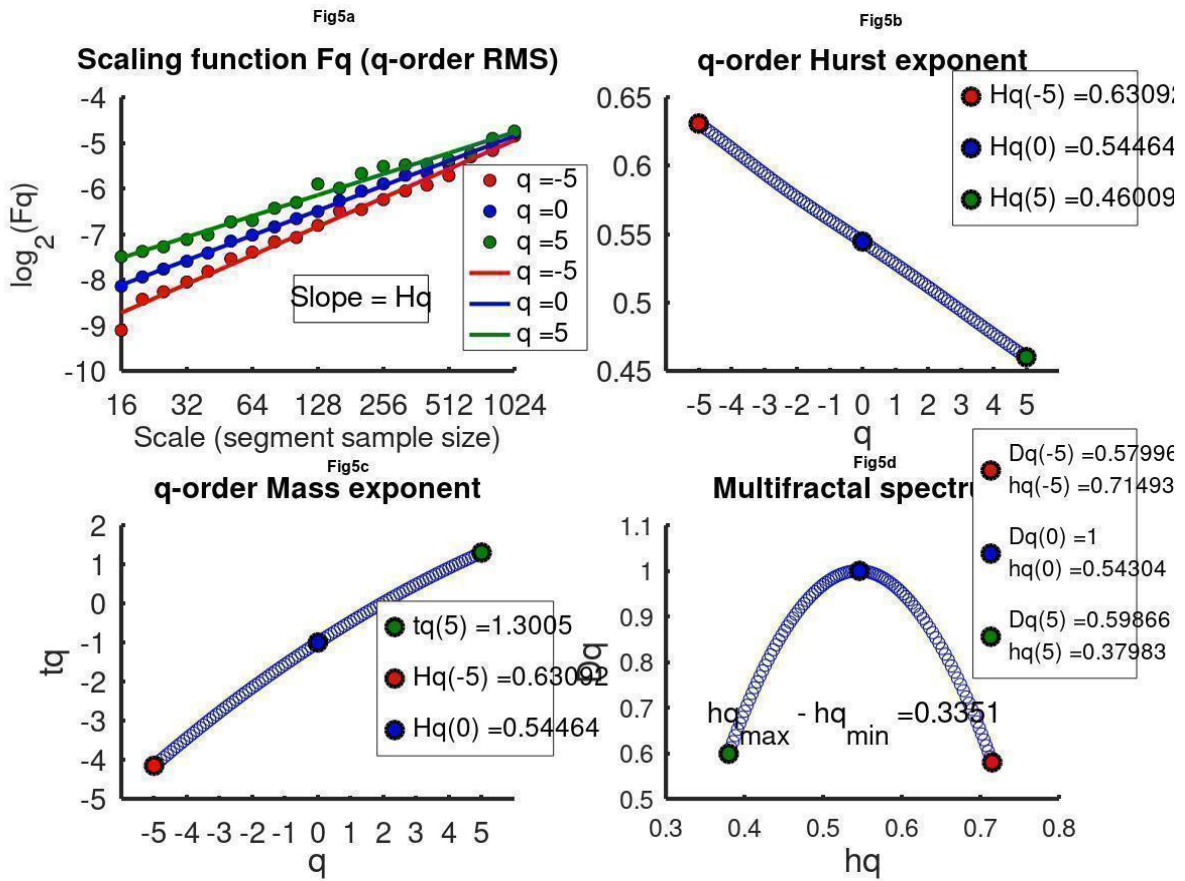

*Figure 5 - The MFDFA calculation results for shuffled USD*

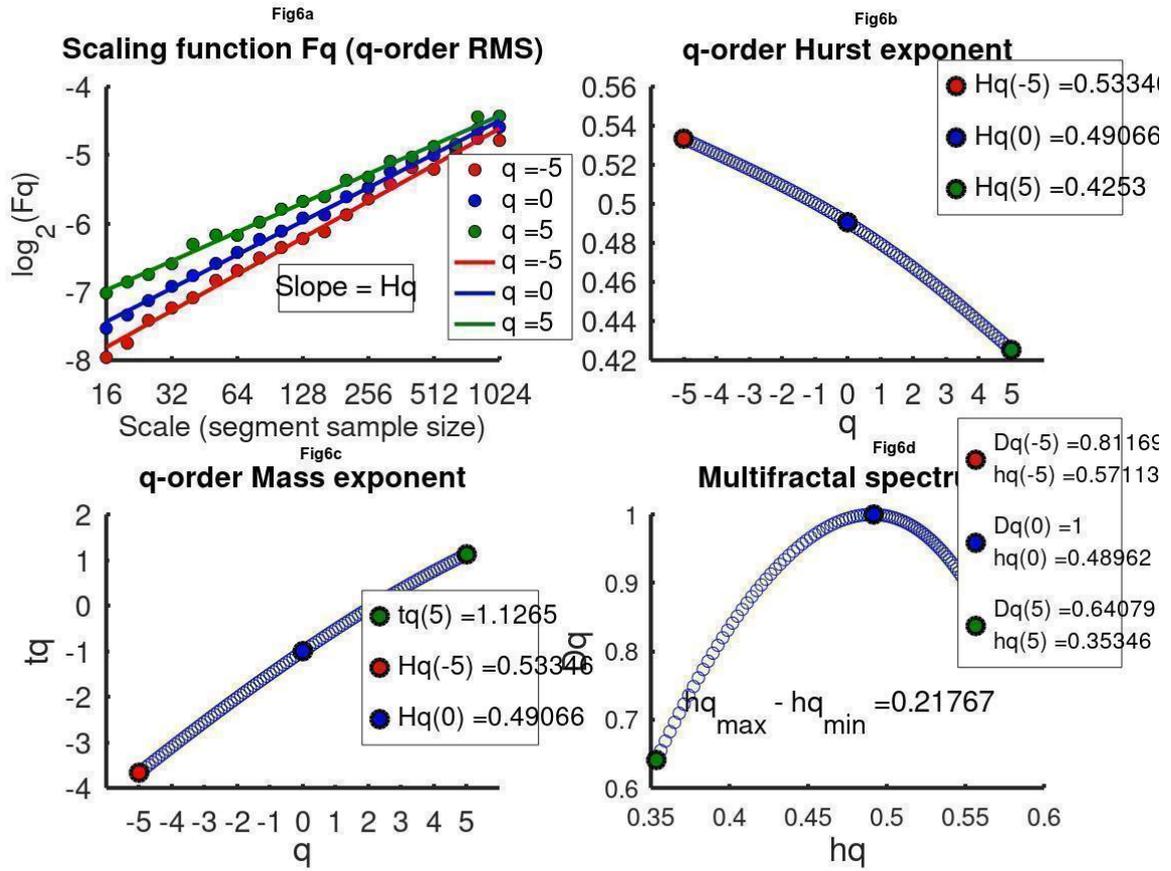

*Figure 6 - The MFDFA calculation results for shuffled GBP*

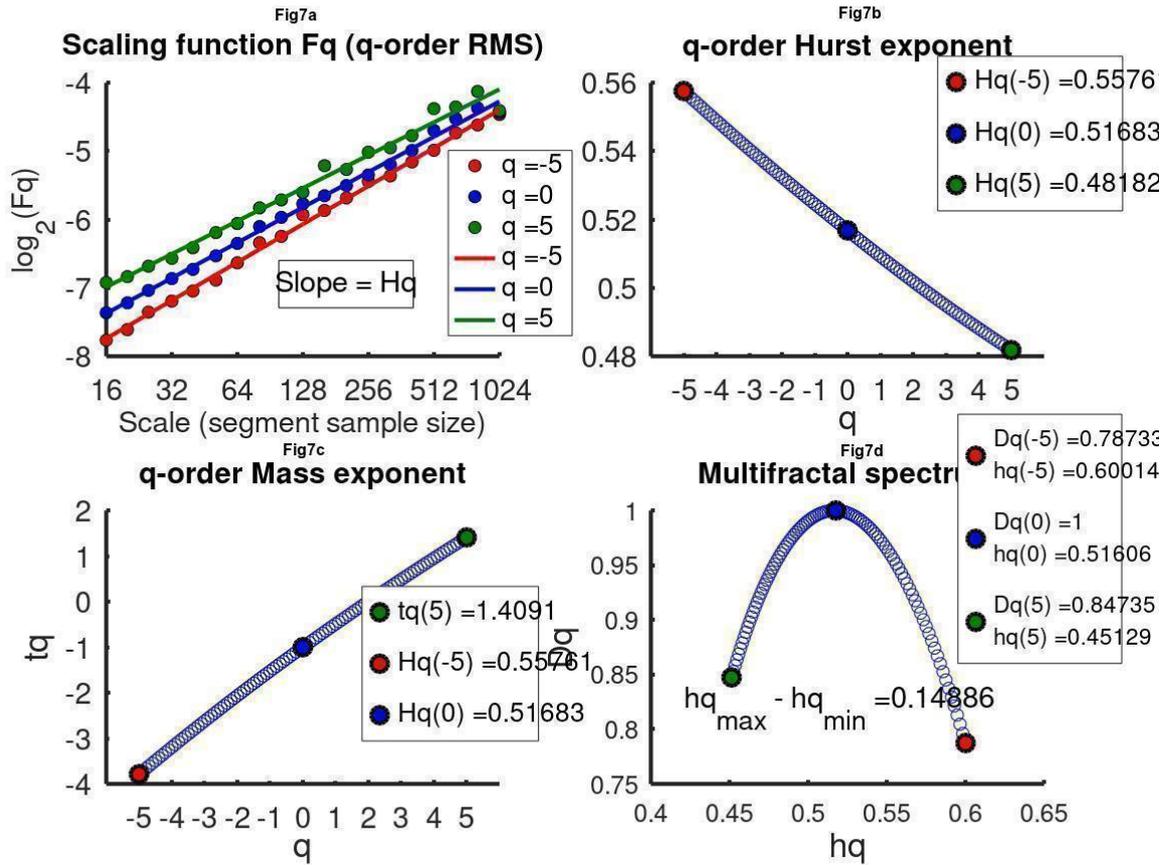

*Figure 7- The MFDFA calculation results for shuffled Euro*

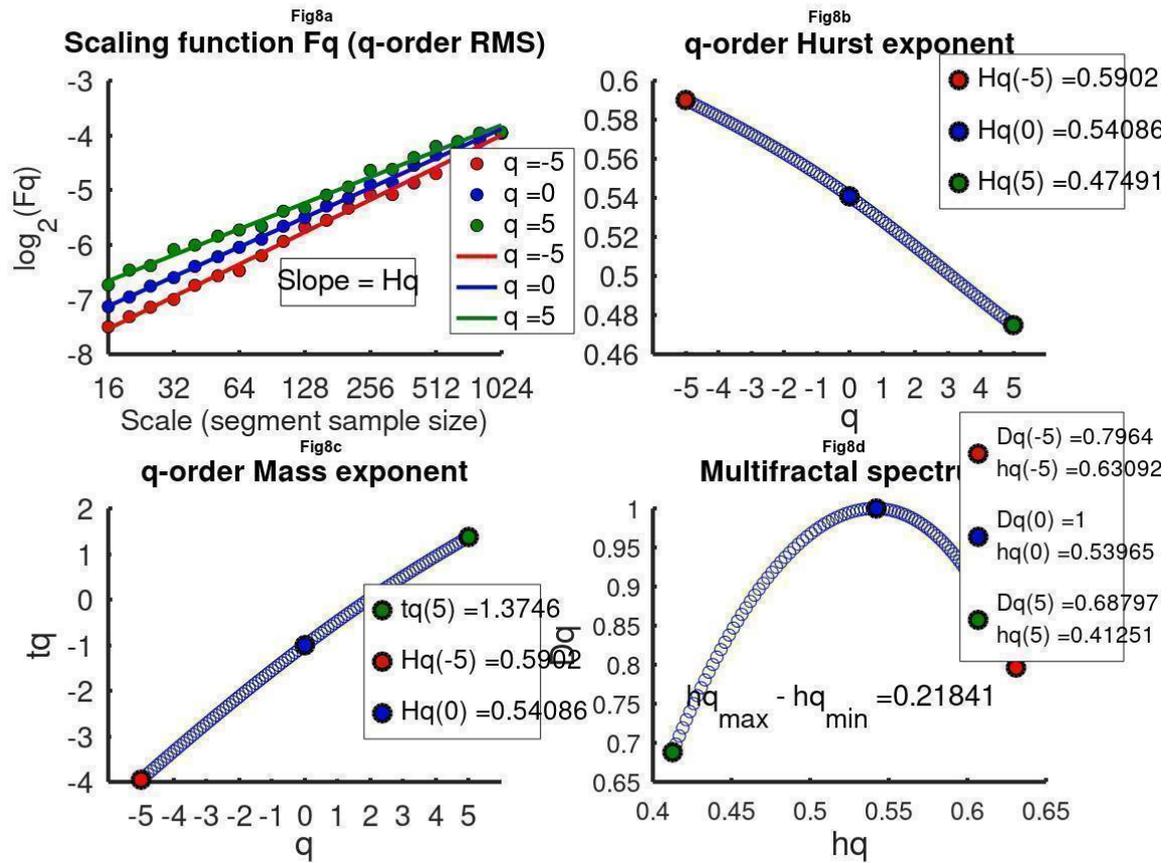

*Figure 8- The MFDFA calculation results for shuffled Yen*

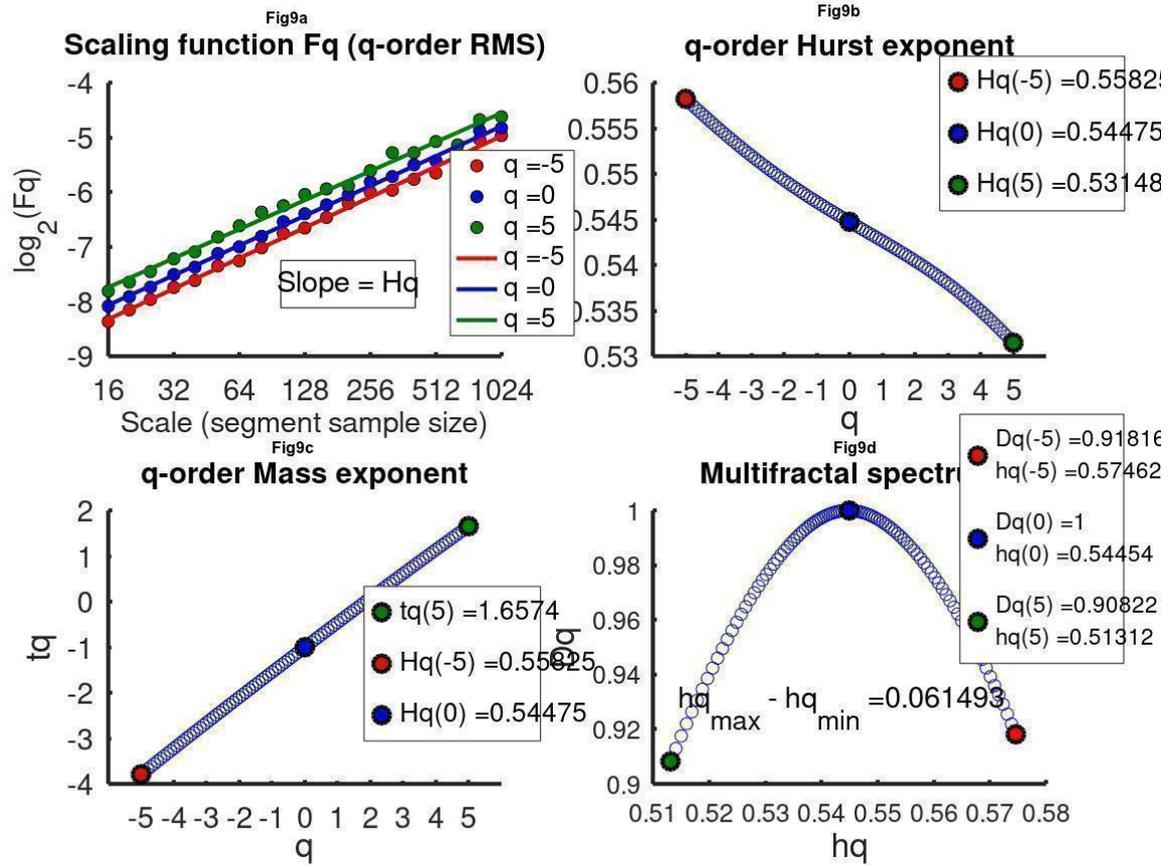

*Figure 9- The MFDFA calculation results for phase randomized USD*

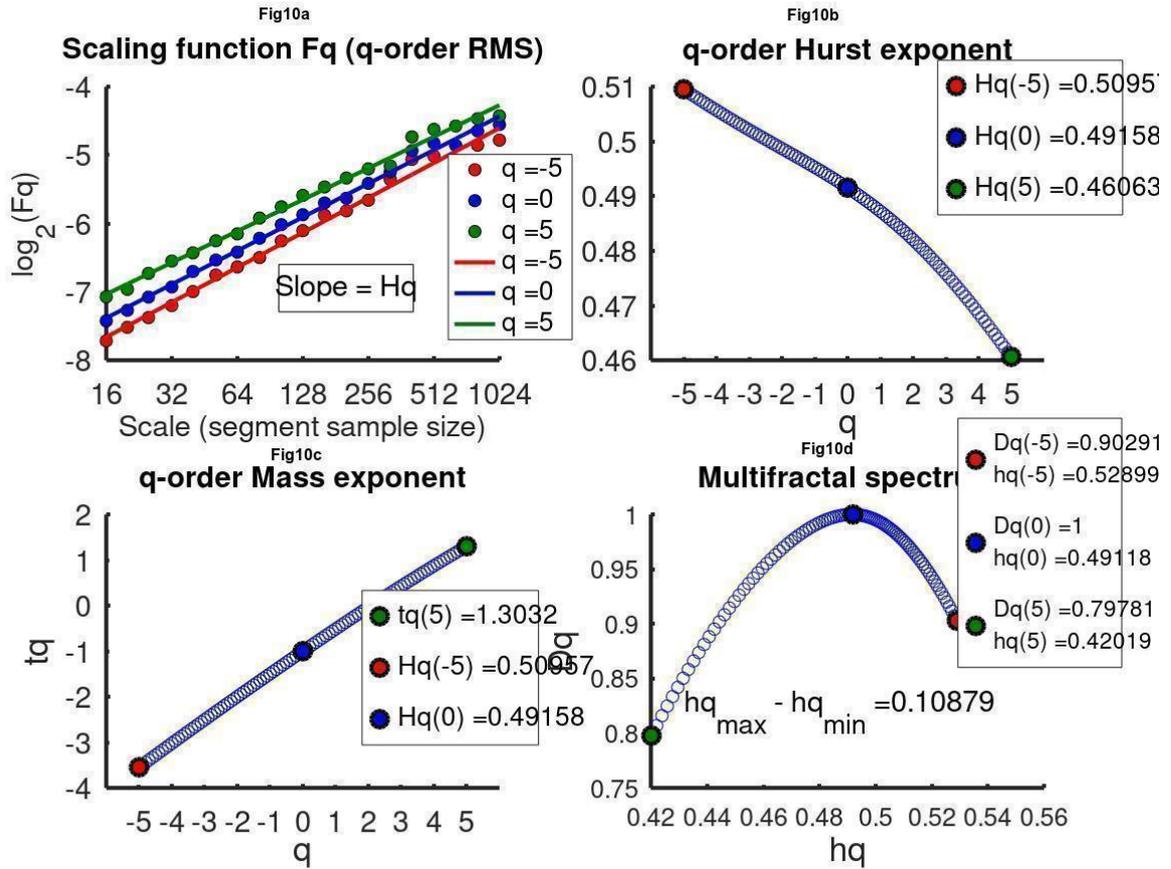

*Figure 10- The MFDFA calculation results for phase randomized GBP*

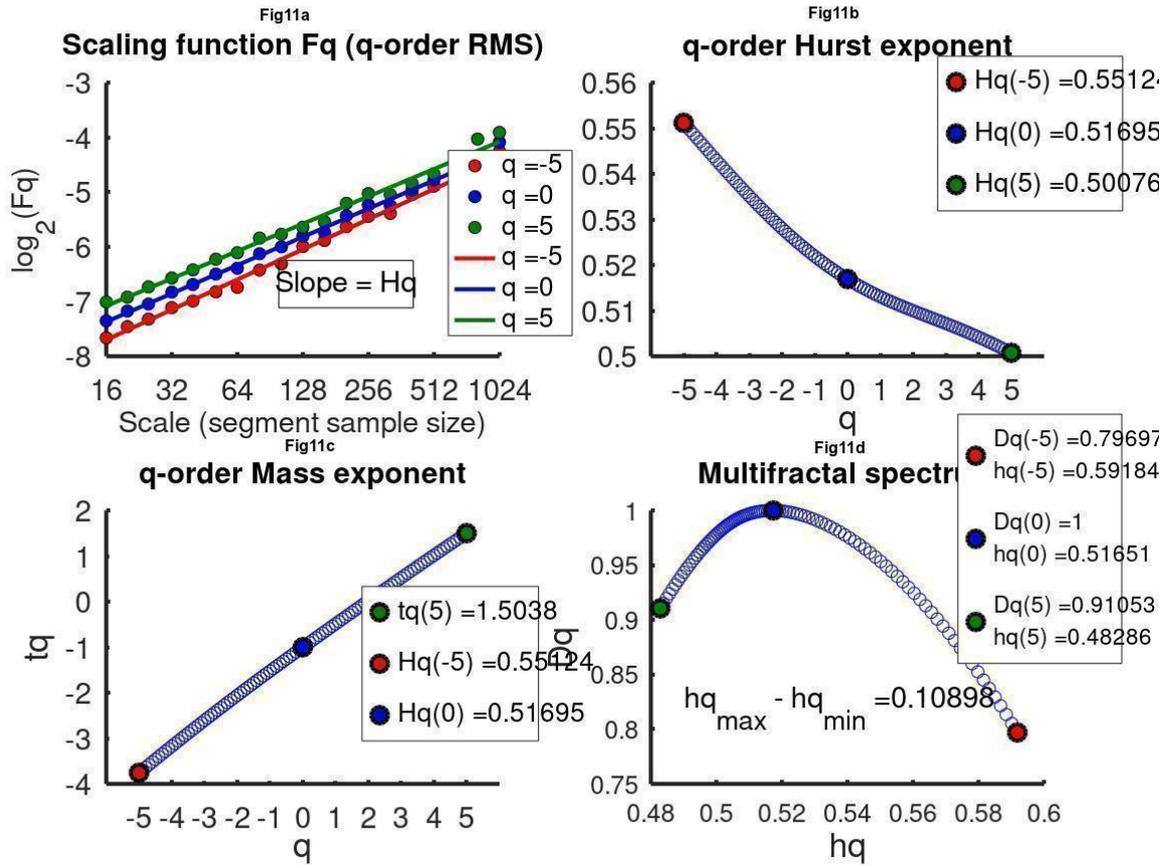

*Figure 11- The MFDFA calculation results for phase randomized Euro*

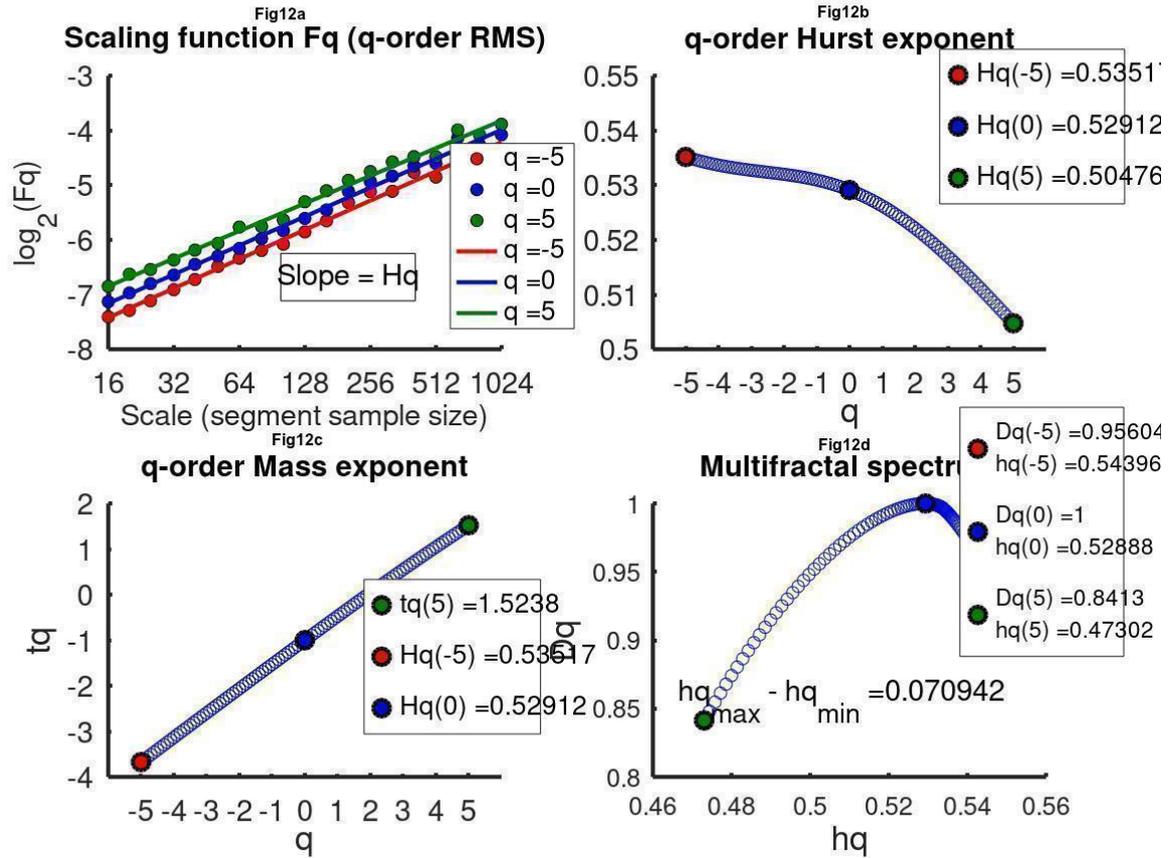

*Figure 12- The MFDFA calculation results for phase randomized Yen*

Table 1 shows the values for the degree of multifractality calculated using $\Delta\alpha = \alpha_{max} - \alpha_{min}$ for the original return time series and for the shuffled time series and the phase randomized time series.

| Foreign Exchange | Original | Shuffled | Phase randomized |
| --- | --- | --- | --- |
|  | $\Delta hq = hq_{max} - hq_{min}$ | $\Delta hq = hq_{max} - hq_{min}$ | $\Delta hq = hq_{max} - hq_{min}$ |
| USD | 0.73166 | 0.3351 | 0.061393 |
| GBP | 0.30326 | 0.21767 | 0.10879 |
| Euro | 0.23831 | 0.14886 | 0.10898 |
| Yen | 0.26528 | 0.21841 | 0.070942 |

Table 1: Strength of Multifractality for original, shuffled and phase randomized series for USD, GBP, Euro and Yen.

It may be observed that as far as the strength of multifractality is concerned the USD has the highest strength followed by the GBP the Yen and the Euro with the least strength.

From Table 1 we can see that for USD the source of multifractality is predominantly due to the fat tail of the distribution since the degree of multifractality is all most removed by phase randomization. There is a decrease in the multifractality strength for the shuffled series signifying some of the multifractality is also attributable to the temporal correlations though not totally.

For the GBP, both the shuffled series and the phase randomized series show a reduction in the strength of multifractality signifying fat tail in the distribution accounting for more multifractality than temporal correlations.

For the Euro, both temporal correlations and fat tail distribution appear to be contributing towards multifractality.

For the Yen, the source of multifractality due to temporal correlations is observed to be much less while the fat tail of the distribution appears to be accounting for the greater part of the multifractality strength.

**5. Conclusion**:

The MFDFA approach has been used in this paper, to research the salient fractal properties of the main Indian exchange rates specifically the US dollar (USD), the British Pound (GBP), the Euro (Euro) and the Japanese Yen (Yen) with respect to the Indian Rupee, from 6th January 1999 to 24th July 2018. We find that the MF-DFA method can explain some of the interesting facets of the return time series of the major Indian forex rates to a significant extent. A nonlinear relationship is observed between the multifractal exponent hq and the order of moment q. A similar nonlinear relationship is also found to exist between the exponent τq and the moment q. The nonlinear variation of hq and τq with respect to the order of moment q reinforces the multifractal nature of the time series studied. Oh et al. (2012) also found evidence of similar multifractal nature in the return series for four ASEAN currencies, the Japanese Yen, the Hong-Kong dollar, the South Korean Won, and the Thai Baht with respect to the US dollar from 1991 to 2005. The degree of multifractality is observed to be highest for the US dollar followed by the GBP and the Yen. The Euro is shown to have the least degree of multifractality. If we relate this to the market efficiency/inefficiency it may be argued that the US dollar market vis a vis the Indian rupee is the least efficient of all the four exchange rates considered in this study while the Euro with the least degree of multifractality may be construed as having the least inefficiency or being the most market efficient. This is of considerable value to foreign exchange traders and intraday speculators who invest in the foreign exchange markets. It may help them to formulate appropriate strategies in designing diversified portfolios with the objective of obtaining superior returns with lower risks.

We also investigate the origins of the multifractality in the return time series by repeating the calculations for the shuffled time series and the phase randomized time series. By analysing the results of these calculations, we find that the multifractality in the return series of the exchange rates are both due to long-range correlations as well as due to broad tails in the distributions, though the nature of this dependency varies from one foreign exchange currency to another. Our results are in agreement with the results found by Kumar and Deo (2009) who have reported about similar relationships while studying the two major Indian stock market indices namely the BSE and the NSE. Our results also seem to agree broadly with the findings of Norouzzadeh (2006) for exchange rate variations for the Iranian Riyal with respect to the US dollar i.e., both studies exhibit the characteristics of multifractality, and the degree is expressed by the widths of the singularity spectra f(α). However, the origin of multifractality of the exchange rate variations data in his case was found to be slightly different in our case. This may be attributed to the different currencies considered as also the period under study.

**Author Declarations**

**Funding -** No funding was received for conducting this study

**Conflicts of interest/Competing interests -** The authors have no competing interests to declare.

**Ethics approval/declarations** - No ethical conflict to declare

**Consent to participate -** Not applicable

**-Consent for publication (include appropriate statements)-** Consent is given for publishing this article in the journal submitted to if it is found suitable by the editor.

**-Availability of data and material/ Data availability (data transparency, if link please provide the link to access. For further information, go to -** Source of data is provided in the paper .

**-Code availability (software application or custom code)**- Not applicable

**-Authors' contributions**  - Dr. Radhika Prosad datta is the sole author of this paper and all the work in this paper was done by him.